\documentclass{emulateapj}
\usepackage{epsf}
\usepackage{epsfig}
\usepackage{graphicx}
\usepackage{xspace}

\newcommand{\ie}{{i.e.}}

\newcommand{\hkpc}{$h_{70}^{-1}$ kpc}

\slugcomment{ }

\shorttitle{Measuring the Geometry of the Universe}
\shortauthors{Taylor et al.}
 
\begin{document}
 
\title{Measuring the Geometry of the Universe\\from Weak Gravitational Lensing behind Galaxy Groups in the HST COSMOS survey}
\author{James E. Taylor\altaffilmark{1},
Richard J. Massey\altaffilmark{2},
Alexie Leauthaud\altaffilmark{3},
Matthew R. George\altaffilmark{4},
Jason Rhodes\altaffilmark{5,6},
Thomas D. Kitching\altaffilmark{2},
Peter Capak\altaffilmark{7},
Richard Ellis\altaffilmark{5},
Alexis Finoguenov\altaffilmark{8,9},
Olivier Ilbert\altaffilmark{10},
Eric Jullo\altaffilmark{6,10},
Jean-Paul Kneib\altaffilmark{10},
Anton M. Koekemoer\altaffilmark{11},
Nick Scoville\altaffilmark{5},
Masayuki Tanaka\altaffilmark{3}}

\submitted{Submitted to ApJ}
\email{taylor@uwaterloo.ca}

\altaffiltext{1}{Department of Physics and Astronomy, University of
  Waterloo, 200 University Avenue West, Waterloo, Ontario, Canada N2L 3G1}

\altaffiltext{2}{Institute for Astronomy, Blackford Hill, Edinburgh
  EH9 3HJ UK}

\altaffiltext{3}{Institute for the Physics and Mathematics of the Universe, The University of Tokyo, 5-1-5 Kashiwanoha, Kashiwa-shi, Chiba 277-8583, Japan}
      
\altaffiltext{4}{Department of Astronomy, University of California,
  Berkeley, CA 94720, USA}

\altaffiltext{5}{California Institute of Technology, MC 249-17, 1200
  East California Boulevard, Pasadena, CA 91125, USA}

\altaffiltext{6}{Jet Propulsion Laboratory, California Institute of Technology, Pasadena, CA 91109}

\altaffiltext{7}{Spitzer Science Center, 314-6 Caltech, 1201
 E. California Blvd. Pasadena, CA, 91125, USA}

\altaffiltext{8}{Max-Planck-Institut f{\"u}r extraterrestrische Physik, Giessenbachstra\ss e,
85748 Garching, Germany}

\altaffiltext{9}{University of Maryland Baltimore County, 1000 Hilltop circle, Baltimore, MD 21250, USA}

\altaffiltext{10}{LAM, CNRS-UNiv Aix-Marseille, 38 rue F. Joliot-Curis, 13013 Marseille, France}

\altaffiltext{11}{Space Telescope Science Institute, 3700 San Martin Drive, Baltimore, MD 21218, USA}

\begin{abstract}
Gravitational lensing can provide pure geometric tests of the structure of space-time, for instance by 
determining empirically the angular diameter distance--redshift relation. This geometric test has been 
demonstrated several times using massive clusters which produce a large lensing signal. In this case, matter at a single redshift dominates the lensing signal, so the analysis is straightforward. It is less clear how weaker signals from multiple sources at different redshifts can be stacked to demonstrate the geometric dependence. We introduce a simple measure of relative shear which for flat cosmologies separates the effect of lens and source positions into multiplicative terms, allowing signals from many different source-lens pairs to be combined. Applying this technique to a sample of groups and low-mass clusters in the COSMOS survey, we detect a clear variation of shear with distance behind the lens. This represents the first detection of the geometric effect using weak lensing by multiple, low-mass systems. The variation of distance with redshift is measured with sufficient precision to constrain the equation of state of the universe under the assumption of flatness, equivalent to a detection of a dark energy component $\Omega_X$ at greater than 99\%\ confidence for an equation-of-state parameter $-2.5 \leq w \leq -0.1$. For the case $w = -1$, we find a value for the cosmological constant density parameter 
$\Omega_\Lambda = 0.85^{+0.044}_{-0.19}$\ (68\%\ C.L.), and detect cosmic acceleration ($q_0 < 0$) at the 98\%\ C.L..  We consider the systematic uncertainties associated with this technique and discuss the prospects for applying it in forthcoming weak-lensing surveys.
\end{abstract}
  
 
\altaffiltext{$\star$}{Based on observations with: the NASA/ESA {\em Hubble 
  Space Telescope} obtained at the Space Telescope Science Institute,
  which is operated by AURA Inc.\ under NASA contract NAS 5-26555; the Subaru
  Telescope, which is operated by the National Astronomical Observatory of
  Japan; the European Southern Observatory under Large Program 175.A-0839,
  Chile; Kitt Peak National Observatory, Cerro Tololo Inter-American
  Observatory, and the National Optical Astronomy Observatory, which is 
  operated by the Association of Universities for Research in Astronomy, Inc. 
  (AURA) under cooperative agreement with the National Science Foundation.}

\section{Introduction}\label{sec:intro}

The current evidence for a dominant dark energy component in the universe \citep[e.g.][]{Percival10, WMAP7, Sullivan11} leaves cosmology in a uncomfortable situation. Given the concurrent evidence for cold dark matter (CDM) and an additional field driving inflation, it seems several radical new components of physics are required to explain the present-day state of the universe, with little detailed observational information so far as to their precise nature. Clarifying the nature of dark energy is particularly challenging. Observationally, the effect of dark energy on the equation of state is very close to that of a cosmological constant $\Lambda$. To determine anything else about this component requires very precise tests and a rigorous elimination of systematics. 

The simplest evidence for dark energy comes from measurements of the geometry of space-time, or equivalently distance as a function of redshift, either from supernovae \citep{Sullivan11}, which measure luminosity distance, or baryon acoustic oscillations (BAO) \citep{Percival10, Beutler11} or the cosmic microwave background (CMB) \citep{WMAP7}, which measure angular diameter distance. Gravitational lensing also provides tests of cosmology, measuring both the matter distribution and how gravity deflects light on large scales. There has been much emphasis on cosmological lensing tests using cosmic shear \citep[e.g.][]{Massey:2007a, Fu, Schrabback:2010, Semboloni}, which is sensitive both to the matter distribution and to space-time geometry. While the theory of CDM structure formation makes fairly robust predictions as to the matter distribution, and thus the lensing potential, on scales larger than galaxies, it also assumes additional physics associated with inflation, such as Gaussian distribution of initial fluctuations with an almost-scale invariant power spectrum. By separating out the geometric effects of lensing from the properties of the density field, one can in principle derive more general constraints on geometry, independent of these assumptions.

The underlying idea of geometric lensing tests is straightforward. First, the strength of lensing is measured behind an object as a function of redshift. Then, taking ratios of the lensing signal at different redshifts, the dependence on redshift gives a measure of space-time geometry via the angular diameter distance--redshift relation. The attraction of this technique is that (at least in principle) any uncertainties in the exact form of the lens potential cancel out. A cosmological constant (or dark energy with a similar equation of state) manifests itself by increasing the distance to an object at a given redshift. In geometric lensing tests, the amplitude of the lensing distortion provides an independent estimate of the distance that the source lies behind the lens. Thus by measuring average distortion versus offset in redshift, one can constrain the value of $\Lambda$ or more generally $\Omega_X$, the dark energy density parameter.

In practice, the specific implementations of this idea fall into two broad categories. The most developed tests use one or a few massive clusters to provide the lensing signal, and measure the position of strongly lensed arcs to determine the strength of lensing as a function of redshift behind the cluster. This method was first discussed in detail by \cite{LinkPierce} (although the idea is much older, e.g.~\citealt{Refsdal66}), who assumed the simplest singular isothermal sphere (SIS) potential for the cluster. The method was subsequently revisited by \citet{Golse02},  who considered several sources of systematic error, in particular  the effect of substructure and irregular cluster mass distributions.  They applied the test to the galaxy clusters AC114 and Abell~2218, obtaining constraints $0 < \Omega_M < 0.33$ on the matter density parameter and $w < -0.85$ on the dark energy equation-of-state parameter \citep{Soucail04}. The method was also used by \citet{Sereno02} on the cluster CL~0024+1654, providing evidence for a flat accelerating cosmology. Most recently, \citet{Jullo10} have derived tight constraints using 28 images from 12 multiple image families in Abell~1689. 

A related method uses weak lensing to measure the amplitude of the lensing signal as a function of redshift behind a cluster \citep[see][ for early forms of this test]{Lombardi99, Gautret}. \citet{Wittman01} first used the weak lensing signal to determine the redshift to a cluster by lensing alone, and it has subsequently been used by \citet{GavazziSoucail07} to estimate redshifts to a dozen clusters in the CFHTLS Deep fields.
Most recently \citet{Medezinski11}, have demonstrated the effect for 3 massive clusters using rough photometric redshifts (photo-zs) to determine mean redshifts to different samples of lensed galaxies.  This work should provide interesting constraints on dark energy when extended to their full sample of 25 clusters.

A second category of tests uses the combined signal from many less massive halos as the source of the lensing effect. \citet{JainTaylor03} proposed the first such test, the `cross-correlation tomography' method. Here the mass distribution is inferred statistically from the foreground galaxy distribution, while the lensing signal is measured with weak shear in two background samples; ratios of the galaxy-shear cross-correlation functions for the two samples then probe geometry. Variants of this method have been developed by \citet{BernsteinJain} and  \citet{Zhang05}. \citet{Taylor07} proposed applying this technique behind clusters using ratios of individual shear measurements, rather than correlation functions. Their revised method was demonstrated together with 3D cosmic shear in \citet{Kitching07} and systematics and error forecasts were discussed in detail in \cite{Kitching08}.

The two sorts of tests have different advantages and disadvantages. Tests using a small number of massive clusters benefit from significant lensing signal and external constraints on the form of the mass distribution from optical or X-ray data, but suffer from cosmic sample variance and uncertainties in the cluster profile. Since these tests probe only one or a few lines of sight, real structures in front of or behind the cluster -- voids or other halos -- will enhance or reduce the signal at particular redshifts. Although ideas for modeling the line of sight are emerging \citep[e.g.][]{Hoekstra11}, the results are not yet satisfactory, and the current practical consensus is to average the signal coming from different clusters. Simulations suggest that at least $\sim$10 massive clusters with many multiple-image systems each are needed to overcome cosmic sample variance \citep{Dalal05,Gilmore09,Aloisio}. Furthermore, massive clusters are relatively rare, so only a fraction of the total lensing signal from all cosmic structure can be used. Tests using a larger number of less powerful lenses sample more of the total lensing signal but require large survey areas with accurate photometric redshifts, and may be subject to more subtle uncertainties in the foreground mass distribution. Furthermore, since the geometric term in the lensing equation depends on source and lens redshifts separately, it is not immediately clear how to stack the results from large samples of lenses in a simple way.  The analysis is thus less intuitive, making it harder to spot unanticipated systematics in the results.

The COSMOS survey provides an interesting data set with which to explore geometric lensing tests. COSMOS has an unusual combination of a high density of sources with accurate lensing shape measurements, and accurate photo-zs for a large fraction of these sources. Unfortunately the COSMOS field has no really massive clusters in it; the largest cluster has an estimated mass of $2.5\times 10^{14} M_\odot$ \citep{Finoguenov:2007}, 10 times less than the largest strong-lensing clusters, and is at a redshift of $z = 0.73$ \citep{Guzzo07} where lensing is past the peak in sensitivity, given the source redshift distribution. The lensing signal in the COSMOS field comes instead from many low-significance, group-sized haloes \citep{Finoguenov:2007,Leauthaud:2010}. Collectively these systems could still produce a large lensing signal to test geometry, however, provided the signal could be stacked. 

Here we propose a simple method for stacking the signal from multiple lenses into a single measure of geometry, in effect the relation between comoving or angular diameter distance and redshift. Applying this new `stacked shear ratio test' to X-ray selected groups in the COSMOS field, we obtain a clear detection of the geometric signal and derive significant constraints on the dark energy density parameter $\Omega_X$. While the COSMOS field is probably too small to overcome sample variance limitations, the magnitude of our statistical errors illustrates the future promise of this technique. 

The outline of paper is as follows: in section 2 we present the basic data, including the sample of lensing groups, and the source redshifts and shape measurements of the COSMOS lensing catalogue. In section 3 we introduce the stacking technique and discuss optimal weighting for this technique. In section 4, we use the stacked shear ratio test to derive parameter constraints on the density of dark energy $\Omega_X$ and the equation-of-state parameter $w$, and discuss possible systematics for this test. In section 5 we summarize our results and discuss future prospects for applying the stacked shear ratio test to other weak lensing surveys. Throughout the paper we refer our results to the WMAP 7-year mean  parameter values derived in \citet{WMAP7}, taking a flat cosmology with $\Omega_\Lambda = 0.73$, $\Omega_M = 0.27$, $H_0 = 70\,h_{70}$ km\,s$^{-1}$\,Mpc$^{-1}$ (WMAP7 hereafter) as our baseline.

\section{Data}\label{sec:data}

\subsection{The COSMOS survey and group sample}

The COSMOS survey \citep{Scoville:2007a} brings together panchromatic imaging
from X-ray to radio wavelengths, including the largest contiguous area
observed with the Hubble Space Telescope, and deep optical spectroscopic
observations. The field covers an area of 1.64 square degrees centered at
10:00:28.6, +02:12:21.0 (J2000) and contains identified groups, clusters and 
larger structures spanning a wide range in redshift \citep{Scoville:2007c}.

We consider the gravitational lensing signal behind a sample of galaxy groups selected originally 
via their X-ray emission \citep{Finoguenov:2007} and updated using a combined mosaic of imaging 
from {\sl XMM-Newton} \citep[1.5 Ms,][]{Hasinger:2007,Cappelluti:2009} and the {\sl Chandra} 
observatories \citep[1.8 Ms,][]{Elvis:2009}. Groups are detected from the combined X-ray mosaic using 
a wavelet filter, which can result in centering uncertainties of up to 32$\arcsec$. The distribution of galaxies along the line of sight to each X-ray detection is searched for a red sequence overdensity to determine the group redshift, with spectroscopic redshifts used for subsequent refinement 
\citep{Finoguenov:2007}.
Group members are selected based on their photometric redshift and proximity to the X-ray centroid, using an algorithm tested extensively on mock catalogs and spectroscopic subsamples \citep{George:2011}. Stellar masses of the member galaxies are determined from multiwavelength data (see \citealt{Leauthaud:2011} for details). From an initial list of members, group centers are then redefined around the most massive group galaxy within the NFW scale radius of the X-ray centroid ({\tt MMGG}$_{\rm scale}$), which optimizes the weak lensing signal at small radii (George et al. in prep.). For the majority of our groups this gives centers which agree with the X-ray centroid; a minority (approximately 20\%) of groups show significant offsets between the most massive galaxy and the X-ray centroid. These offsets could be due to observational problems (such as low signal-to-noise in the X-ray or optical data), or they might indicate unrelaxed, low-concentration groups with poorly defined physical centers, such as recent mergers. We will consider below both the full set of groups and a `restricted' set which excludes the systems with significant offsets. The centering algorithm will be discussed further in a forthcoming paper (George et al.~in prep.). The full X-ray group sample, together with derived properties, will be made available through the NASA/IPAC Infrared Science Archive (IRSA) website\footnote[1]{http://irsa.ipac.caltech.edu/Missions/cosmos.html} (see \citealt{George:2011} for details).

\begin{figure}[t]
\includegraphics[width=0.48\textwidth, keepaspectratio]{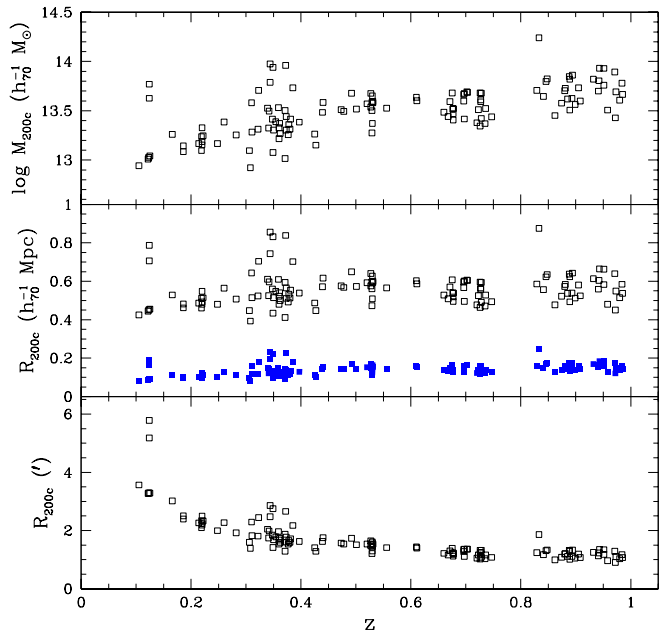}
 \caption{Top panel: group mass versus redshift. Middle panel: estimated virial radius in $h^{-1}_{70}$\,Mpc (filled circles). The solid squares show the estimated scale radius, which is typically 1/4--1/5 of the virial radius. Bottom panel: estimated angular extent of the virial radius. In each case, only the 129 groups with well-determined centers and redshifts are shown, as described in section 2.1. A WMAP7 cosmology is assumed.}
 \label{fig:xgroups}
\end{figure}

We restrict the lens sample to groups at $z < 1$ to ensure the reliability of X-ray detections and optical associations, as well as good photometric redshifts for identifying members and centers. We further cut out of the sample poor groups, groups with centroids affected by masking, and possible mergers (this 
corresponds to taking only groups with {\tt FLAG\_INCLUDE} = 1 as defined in \citealt{George:2011}). Our final sample consists of 129 systems (105 in the restricted set) spanning a rest-frame 0.1--2.4~keV luminosity range between $10^{41}$ and $10^{44}$erg/s, with estimated virial masses of 
$0.8\times10^{13}$ -- $2\times10^{14}\, h^{-1}_{70}$\,M$_\odot$, virial radii of 0.4--0.8\,$h^{-1}_{70}$Mpc, 
and projected angular sizes of $1\arcmin$--$6\arcmin$. Fig.~\ref{fig:xgroups} shows the mass, physical size and angular size for the groups in the sample (note units have been converted from the value $H_0 = 72$ km s$^{-1}$ Mpc$^{-1}$ used in the catalog to $H_0 = 70$ km s$^{-1}$ Mpc$^{-1}$). The virial radius is taken to be $R_{200c}$, the radius within which the mean density is equal to 200 times the critical density  $\rho_c(z)$ at the redshift of the group, and the virial mass is taken 
to be  $M_{200c}$, the mass enclosed within $R_{200c}$. These masses and radii are estimated from the X-ray data, using X-ray scaling relations calibrated with lensing data \citep{Finoguenov:2007, Leauthaud:2010, George:2011}.

\subsection{Weak lensing galaxy shape measurements}\label{subsec:shapes}

High resolution imaging of the COSMOS field was obtained with the Hubble Space
Telescope between October 2003 and June 2005 \citep{Scoville:2007b,Koekemoer:2007}. 
The main program consisted of 575 slightly overlapping pointings of the
{\it Advanced Camera for Surveys} (ACS) {\it Wide Field Camera} (WFC) taken with the
$F814W$ (approximately $I$-band) filter. At each pointing, four slightly dithered
exposures of 507~seconds were obtained. Any cosmetic defects and reflection ghosts
were carefully masked by hand. Using Version 2.5.0 of the SExtractor
photometry package \citep{Bertin:1996}, in a \textit{Hot-Cold} configuration on
the stacked images, we detected compact objects in a $0.15\arcsec$ diameter
aperture down to $F814W_{AB}=26.6$ at $5\sigma$ \citep{Leauthaud:2007}.

We measure the shapes of galaxies in this catalogue using the RRG method \citep{Rhodes:2000},
largely following the analysis pipeline of \citet{Leauthaud:2007} and calibrated
against simulated ACS images generated with the {\tt simage} package
\citep{Massey:2004,Massey:2007b}. However, we now include two significant
improvements on this earlier work.

First, we correct trailing in the ACS images due to charge transfer inefficiency (CTI) via a
physically-motivated readout model \citep{Massey:2010} that acts at the pixel
level, rather than a parametric scheme at the catalog level. This moves electrons
in the raw exposures back to where they should have been read out. The method achieves a
97$\%$ level of correction and is robust to variety in galaxy morphology,
local galaxy density and sky background level. After correction, residual shears 
are well below statistical measurement precision.

Second, we model the Hubble Space Telescope's point spread function (PSF) as a function
of chip position, telescope focus offset, and velocity aberration factor
(VAFACTOR). The latter two parameters reflect HST's thermal condition during each
exposure: slight expansion and contraction changes the PSF. Following
\citet{Rhodes:2007}, we measure the focus offset (the distance between the
primary and secondary mirrors) with a precision of $1\mu$m by comparing the
apparent shapes of $\sim10$ stars in each exposure to {\tt TinyTim} models
\citep{Krist:2003}. \citet{Jee:2007} found that focus offset correlates with the
first Principal Component of PSF variation and accounts for 97\%\ of the power, while
\citet{Schrabback:2010} found that VAFOCUS correlates with the next Principal
Component. We measure the shapes of all stars in the COSMOS imaging, then
interpolate between them using all four measured parameters. This improves residuals compared to \citet{Massey:2007a}, and we retain this physically-motivated approach rather than relying solely on Principal Component Analysis. 

We have also revised our method for determining the variance of the tangential shear slightly. This is now determined empirically, as described in section 3.5 of \citet{Leauthaud:2011}. Galaxies are binned by S/N and magnitude, and the total variance of the shear components $\gamma_1$ and $\gamma_2$ is measured directly in each bin. This empirical derivation of the shear dispersion includes both the scatter due to intrinsic variations in galaxy shape, and the additional scatter due to shape measurement errors. We find that the shear dispersion varies from 
$\sigma_{\tilde{\gamma}} \sim 0.25$ for bright galaxies with high S/N to 
$\sigma_{\tilde{\gamma}} \sim 0.4$ for faint galaxies with low S/N. These measured values may be very slightly overestimated, however, as suggested by the reduced $\chi^2$ of the profile fit discussed 
in section \ref{subsec:radial} below.

\subsection{Photometric redshift measurements}

Of the 129 groups in our full sample, 95\%\ contain two or more spectroscopically confirmed members, 3\%\ have one spectroscopically confirmed member, and the remainder have redshifts determined photometrically from the red sequence of member galaxies \citep[as in][]{Finoguenov:2010}. The average redshift error for the group ensemble is $\sim0.0017$, only slightly larger than their typical velocity dispersions of 300--450 km/s.

Multicolor ground-based imaging in over 30 bands \citep{Capak:2007} also
provides photometric redshift information for all of the source galaxies along the 
same line of sight. We use photometric redshift estimates from the {\tt LePhare} $\chi^2$
template-fitting code, which are updated from those published in \citet{Ilbert:2009} by the
addition of deep $H$-band data, and small improvements in the template fitting
technique. We have compared photo-$z$s from 10801 galaxies at $z\sim0.48$, 696 at
$z\sim 0.74$, and 870 at $z\sim 2.2$ to spectroscopic redshift measurements with
the Very Large Telescope (VLT) Visible Multi-Object Spectrograph (VIMOS)
\citep{Lilly:2007} and the Keck Deep Extragalactic Imaging Multi-Object
Spectrograph (DEIMOS). The r.m.s.~dispersion in the offset $\sigma_{\Delta z}$
between photometric and spectroscopic redshift is $0.007(1+z)$ at
$i^{+}_{AB}<22.5$ and $0.02(1+z)$ at $i^{+}_{AB} \sim 24$ 
and $z<1.3$ (or $0.06(1+z)$ for $i^{+}_{AB} \sim 24$ and $z\ge 1.3$).

To mitigate against catastrophic failure in estimated photo-$z$s, for example due
to confusion between the Lyman and $4000$\AA\  breaks, we reject from the sample
all source galaxies with a secondary peak in the redshift probability distribution 
function (\ie\ galaxies where the parameter {\sc zp\_sec} is
greater than zero in the \citealt{Ilbert:2009} catalog). The rejected {\sc zp\_sec $>0$} 
galaxy population is expected to contain a large fraction of
catastrophic errors \citep[roughly 40\%--50\%\ --][]{Ilbert:2006,Ilbert:2009}. 
For the purposes of cosmological constraints, we further exclude from the sample
objects with relative redshift uncertainties $\Delta z/(1+ z) \ge 0.05$, taking the average 
redshift error to be $\Delta z \equiv ({\text{\tt zu68\_gal}} - ${\tt zl68\_gal}$)/2.0$, where {\tt zu68\_gal} 
and {\tt zl68\_gal} are the 68\%\ confidence limits on the redshift, based on the 
photo-z probability distribution \citep{Ilbert:2009}.
Our final source sample consists of all galaxies passing these cuts that lie within 
6$\arcmin$ of a group center. Individual galaxies may enter into the final 
sample multiple times if they lie within 6$\arcmin$ of more than one peak. 
The photo-z quality cuts reduce the number density of source galaxies to $26$ 
galaxies/arcminute$^{2}$, for a total of $3.7\times 10^5$ galaxies 
($3.1\times 10^5$ in the restricted sample). The mean redshift of the final sample is 
$\langle z \rangle = 0.95$ and the mean relative error in redshift is  
$\Delta z/(1+ z) = 0.018$, while the mean magnitude is $\langle \rm{I}_{F814W} \rangle \sim 24$.  

\section{Method}\label{sec:results}

\subsection{The Stacking Method}\label{subsec:stacking}

If we consider a source galaxy (or `source' hereafter) at redshift $z_S$ being lensed by a foreground group (or `lens' hereafter) at redshift $z_L$ and observed at $z_O$, in the weak limit the tangential shear induced by the lens will be:
\begin{equation}
\gamma_t(r) = \left[\, \overline{\Sigma}(<r) - \overline{\Sigma}(r)\right]/\Sigma_c = \Delta\Sigma(r)/\Sigma_c
\label{eqn:basicshear}
\end{equation}
where $\overline{\Sigma}(<r)$ is the mean surface density interior to projected (physical) radius $r$, $\overline\Sigma(r)$ is the azimuthally averaged surface density at $r$, and $\Sigma_c$ is the critical surface density defined as:
\begin{equation}
\Sigma_c \equiv {{c^2}\over{4\pi G}}{{D_S}\over{D_L D_{LS}}}\, .
\end{equation}
Here $D_S$ denotes the angular diameter distance from the observer to the source
\begin{equation}
D_S = f_k(\omega[z_O,z_S])a(z_S)\, ,
\end{equation}
where $\omega[z_O,z_S]$ is the comoving (or coordinate) distance along a radial ray between the observer and the source
\begin{equation}
\omega[z_O,z_S] = \int_{z_O}^{z_S} dw = \int_{a_S}^{a_O} {{cdt}\over{a(t)}} = \int_{a_S}^{a_O} {{cda}\over{a^2H(a)}}\, .      
\end{equation}
$D_L$ and $D_{LS}$ are angular diameter distances from the observer to the lens and from the lens to the source respectively, given by 
\begin{eqnarray}
D_L &=& f_k(\omega[z_O,z_L])a(z_L)\,,\\
D_{LS} &=& f_k(\omega[z_L,z_S])a(z_S)\,, 
\end{eqnarray}
\citep[e.g.][]{BS01}.

If we consider the case of a flat cosmology ($k = 0$), $f_k(\omega) = \omega$ and thus we can rewrite $\Sigma_c$ more simply in terms of the comoving distances $\omega_L \equiv \omega[z_O,z_L]$ and $\omega_S \equiv \omega[z_O,z_S]$:
\begin{equation}
\Sigma_c \equiv {{c^2}\over{4\pi G}}{{\omega_S}\over{\omega_L (\omega_S - \omega_L)}} (1 + z_L).
\end{equation}
The critical density incorporates all the geometric dependence of lensing; it is the non-linear dependence  of $\Sigma_c$ on $z_S$ and $z_L$ that makes it hard to stack the signal from different source-lens pairs in a straightforward way. We can simplify the dependence, however, by defining the comoving distance ratio $x \equiv \omega_S/\omega_L$. Note that $x > 1$ for sources chosen to be behind the lens; we will consider objects with both $x > 1$ and $x \le 1$ below. In terms of $x$,
\begin{equation}
\Sigma_c \equiv {{c^2}\over{4\pi G}}{{(1 + z_L)}\over{\omega_L}}{{1}\over{(1 - 1/x)}}\, . 
\end{equation}
We can also write this in terms of $\Sigma_{c,\infty}$, the value of the critical density in the limit $x \rightarrow \infty$ :
\begin{equation}
\Sigma_c = {{1}\over{(1 - 1/x)}}\Sigma_{c,\infty}\,, 
\end{equation}
where 
\begin{equation}
\Sigma_{c,\infty} \equiv {{c^2}\over{4\pi G}}{{(1 + z_L)}\over{\omega_L}} 
\end{equation}
depends only on the lens properties, not on the properties of the source galaxy.

From Eqn.~\ref{eqn:basicshear},
the geometry of all source-lens pairs now takes on a universal form
\begin{equation}
\gamma_t(r)\Sigma_{c,\infty}/\Delta\Sigma(r) \equiv \Gamma(x) = \left(1 - {1\over{x}}\right)\, .
\end{equation}
$\Gamma(x)$ corresponds, e.g., to the lensing efficiency $E$ defined by \citet{Golse02}.
In as much as the measured tangential ellipticity $\varepsilon_t$ of each source galaxy is an estimator $\tilde{\gamma}_t$ of the true tangential shear $\gamma_t$, we can construct a weighted sum of estimates from individual source galaxies $j$ with respect to lensing centers $i$ to recover the universal geometric dependence: 
\begin{eqnarray}
\left(1 - {1\over{x}}\right) &=& \sum_{i,j} w_{ij}\Gamma_{ij}(x) \nonumber\\ 
&=& {{\sum_{i,j} w_{ij}\tilde{\gamma}_{t,ij}\Sigma_{(c,\infty) i}/\Delta\Sigma_i(r_{ij})}\over{\sum_{i,j} w_{ij}}}
\label{eqn:gamma}
\end{eqnarray}
with weights $w_{ij}$ chosen to maximize the signal-to-noise ratio or sensitivity to cosmological parameters, as discussed below. 

Since we are just fitting the data to a fixed function, cosmology appears to have disappeared from Eqn.~(\ref{eqn:gamma}). In fact, it is hidden in the conversion from measured source and lens redshifts to inferred source and lens distances. For a given cosmology we convert redshifts to comoving or angular diameter distances, construct the weighted sum in Eqn.~\ref{eqn:gamma}, 
and calculate the $\chi^2$ of $\Gamma(x)$ with respect to the theoretical expectation $(1 - 1/x)$. This gives us the relative likelihood of that particular set of cosmological parameters; iterating over this process then allows parameter constraints. The only remaining problems are to determine the optimal weights $w_{ij}$ and surface mass density contrast $\Delta\Sigma (r)$. We discuss these calculations in the next section.

Before proceeding we should note that our simple stacking analysis ignores several complications. First, it ignores the distinction between true and reduced shear \citep[e.g.~][]{Shapiro}. In the weak shear limit the two are identical,
and for the groups considered here the surface mass density is low enough that the contribution from non-weak shear corrections is unimportant outside $\sim$ 50--100 \hkpc. The effect of the second-order correction term is illustrated in figure 4 of \citet{Leauthaud:2010}. Its contribution is roughly comparable to that of the stellar mass in the central group galaxy (see Fig.~\ref{fig:radialprofile} below) at large projected radii, and always less than the stellar contribution at radii less than 
$\sim$ 50--100 \hkpc. Given only that $\sim$2\%\ of our sources lie at such small projected radii, 
the effect of these contributions on our fits should be negligible (excluding from the sample all sources within 15$\arcmin$ of group centers\footnote[2]{Note that in order to evaluate $\Delta\chi^2$ smoothly as we vary the cosmological parameters, the sample selection has to be independent of cosmology, and thus our cuts have to be in an observable such as angular separation, rather than a cosmology-dependent quantity such as physical separation.}, for instance, has no effect on the final results). 

Second, we have also assumed flatness in separating the dependence on the lens distance and the source distance. While current cosmological constraints indicate an almost completely flat universe (e.g.~\citealt{WMAP7} find $|\Omega_k| \lesssim 0.01$ from various sets of constraints), it would be nice to be able to relax this assumption. Unfortunately, while it is still possible to fit shear ratios between individual pairs in the general case, there is no simple way of stacking all measurements together into a single functional form, since the dependence on the two redshifts can no longer be factored out of 
$f_k(\omega[z_L,z_S])$ in a simple way. We can estimate the effect of curvature by considering the series expansion for $f_k$ in the limit $|\Omega_k| \ll 1$. For typical values of $\omega \sim c/H_0$, 
the next term in the series is smaller by a factor $1/6 (\omega/R_0)^2 \sim 1.6\times 10^{-3}$, where 
$R_0 \equiv c/(H_0 \sqrt{|\Omega_k|})$. Thus in realistic non-flat cosmologies, we expect a correction of order $1.6 \times 10^{-3}$ to our values of $\Gamma(x)$. Compared to our cosmological sensitivity 
$\Delta\Gamma \sim 0.05$ (see section \ref{subsec:paramconstraints} below), this represents a 3\%\ correction to our derived parameters. This correction is smaller than errors on 
$\Omega_X$  we obtain below, although it would be quickly become important in larger surveys. 
In what follows we will ignore the complication of non-zero spatial curvature. 

Third and lastly, our analysis assumes a specific functional form for the surface 
mass density contrast $\Delta\Sigma(r)$, namely a projected NFW profile. 
We will show below that this functional form is in fact an excellent fit to the 
stacked data. We could use instead an empirical profile
based directly on the data itself, but given the agreement 
between the NFW model and the data, this would not 
affect our results significantly.

\subsection{Radial Dependence and Optimal Weighting}\label{subsec:radial}

Our goal is to measure the redshift dependence of the group lensing signal. The radial dependence of the surface mass density contrast $\Delta\Sigma(r)$ around groups, although intrinsically very interesting, is essentially a nuisance parameter in this calculation. We need to determine $\Delta\Sigma(r)$, however, in order to weight measurements of individual source galaxy shapes optimally when estimating $\Gamma(x)$. The density contrast around groups was studied in detail in \citet{Leauthaud:2010}; we reproduce the same calculation here, stacking with respect to physical radius the signal from all groups with well-determined redshifts and centers. 

The density contrast profile includes contributions from four main terms\footnote[3]{
In some cases, a fifth component might be necessary to account 
for the mis-identiÞcation of the central galaxy \citep{Johnston}. 
In our case, we neglect this term which is sub-dominant in 
our sample because we have optimized the centering using weak 
lensing (George et al in prep.).}: the weak shear contribution of the main halo (the `1-halo' term), the average weak shear contribution from nearby halos (the `2-halo' term), a weak shear contribution from the stellar mass of the central group galaxy, and the second-order corrections to the shear in the center of the main halo. Of these, only the 1-halo term is important here; the 2-halo term only becomes significant at large projected radii ($r \sim 4\, h_{70}^{-1}$ Mpc), while the stellar and second-order terms are only significant at small projected radii ($r \lesssim $ 50--100 \hkpc) where we have very few galaxies in the source sample, as discussed in Section \ref{subsec:stacking}.

We expect that the 1-halo term for a single group should follow a projected Navarro-Frenk-White (NFW) profile $\Sigma(r)$, whose form $f_{NFW}(r/r_s)$ is given, e.g.~in \citet{WrightBrainerd00}. The profile has two free parameters, a scale radius $r_s$ and a normalization $\Sigma_0$, or alternately it can be defined in terms of a virial radius $r_{vir}$ and a concentration $c \equiv r_{vir}/r_s$. The expected values of these parameters can be inferred from X-ray fluxes, X-ray scaling relations, and theoretical concentration-mass relations. Using the concentration relations of \citet{Zhao:2009}, for instance, \cite{Leauthaud:2010} predicted concentrations in the range 3.6--4.6 for the COSMOS groups (these values assume the definition $r_{vir} = R_{200c}$; thus they correspond to the values $c_{200c}$ 
 from \citealt{Zhao:2009}).
Simulations suggest that individual halos will have significant ($\sim$50\%) scatter around these mean values \citep{Zhao:2009}. Finally, we note that here we are considering an average profile for all groups in the sample, where the averaging is weighted by surface mass density contrast $\Delta\Sigma$. At a fixed redshift and fixed concentration, $r_s$ and $r_{vir}$ for each group should scale as $M^{1/3}$ and our stacked profile would be similar to a mass-weighted average. Variations in concentration and redshift complicate this behavior, but we can still use mass-weighted averaging to guide our expectations as to the final values for the concentration or scale radius. Using the concentration relations of \citet{Zhao:2009} and assuming a WMAP7 cosmology, for instance, we predicted a mass-weighted average scale radius of $r_s$ = 154 \hkpc\ for our groups. 

\begin{figure}
  \includegraphics[width=0.45\textwidth, keepaspectratio]{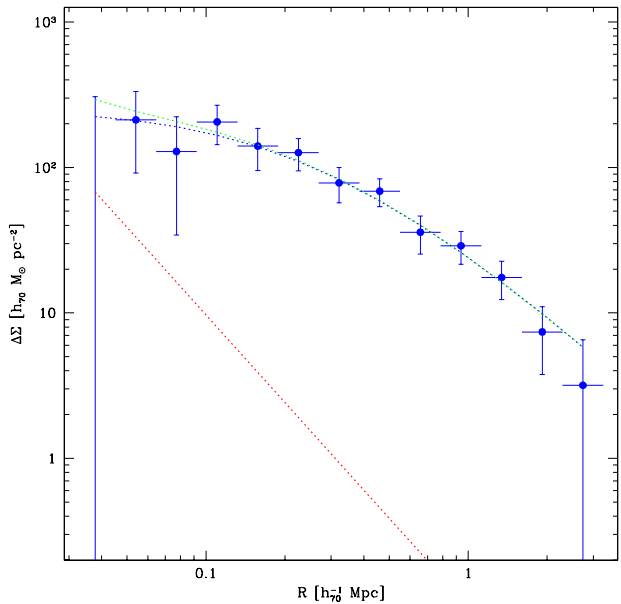}
  \caption{The surface mass density contrast as a function of projected physical separation, stacking all groups with well-defined redshifts and centers (the `restricted' sample). The thin dotted lines indicate contributions from a projected NFW profile (middle blue curve), the average stellar mass of the central galaxy in the group (lower red line), and the sum of these two components (upper green curve). 
}
  \label{fig:radialprofile}
\end{figure}

The surface mass density contrast $\Delta\Sigma(r)$ is related to the tangential shear by
\begin{equation}
\Delta\Sigma(r) =  \Sigma_c \times \gamma_t(r)\, .
\end{equation}
Thus $\Delta\Sigma(r)$ can be estimated as in Eqn.~8 of \citet{Leauthaud:2010}
\begin{equation}
\Delta\Sigma(r) = {{\sum_{ij}w_{ij}{\tilde{\gamma}}_{t,ij}\Sigma_{c,ij}}\over{\sum_{ij}w_{ij}}}\, .
\label{eqn:surfdensityest}
\end{equation}
Fig.~\ref{fig:radialprofile} shows the surface mass density contrast for our stacked sample of groups. 
Points with error bars indicate the mean value inferred from Eqn.~\ref{eqn:surfdensityest}, binned logarithmically in radius. The thin dotted lines indicate contributions from a projected NFW profile (middle blue line), the stellar mass of the central galaxy in the group (lower red line), and the sum of these two components (upper green line). The normalization of the stellar contribution is based on the mean stellar mass of the central group galaxy, as inferred from photometry (see \citealt{Leauthaud:2011} for details).

The main part of the profile is well fit by a projected Navarro-Frenk-White (NFW) profile $\Sigma(r)$ with a scale radius $r_s \sim$ 160--180 \hkpc\ and a normalization $\Sigma_0 \sim 200\, h_{70}$ M$_\odot$ pc$^{-2}$ at $r_s$. The best fit values of these two parameters are strongly correlated, so we choose instead to normalize the profile at a radius $r_{\rm fix}$ = 250 \hkpc\ where the measurement errors are small and the amplitude is less dependent on $r_s$. Thus the profile can be written
\begin{equation}
\Sigma(r) = \Sigma_0 {{f_{NFW}(r/r_s)}\over{f_{NFW}(r_{\rm fix}/r_s)}}\, .
\end{equation}
This gives us constraints on $\Sigma_0 = \Sigma(r_{\rm fix})$ and $r_s$ which are more independent of each other. 

We determine values for $\Sigma_0$ and $r_s$ by calculating $\chi^2$ with respect to a projected NFW model. We do not bin the data in radius, since the bin boundaries would change with cosmology or scaling, but sum the contribution from each galaxy individually. For the full group sample and assuming WMAP7 cosmology, we obtain the lowest reduced $\chi^2$ by considering all sources at $x > 1.1$, with no other cut on redshift errors. This gives a slightly high value for the scale radius $r_s = 187^{+54}_{-29}$ \hkpc, however. Considering only the restricted sample and/or sources with $\Delta z/(1+z) \le 0.05$, we obtain values of $r_s$ closer to the expected value. This suggests that the remaining 20\%\ of the group sample may be affected by centering problems, or that it may include many unrelaxed or disturbed groups with lower mean concentrations. The dependence on photo-z cuts could indicate that the lensing signal in all groups is diluted by group member contamination in the source population when the limits on photo-z errors are relaxed. For the restricted group sample with the cut on source redshift errors, we obtain best fit values $\Sigma_0 = 98.8 \pm 11\,h_{70}$\,M$_\odot$pc$^{-2}$ and $r_s = 158^{+55}_{-28}$ \hkpc. The reduced $\chi^2$ is marginally higher for this sample than for the much larger uncut sample, but the best-fit value of $r_s$ closer to the expected value, so we will take this as our fiducial profile, and marginalize over values of $r_s$ and $\Sigma_0$ in this range for our cosmological calculations. The best fit value of $r_s$ also places some constraints on possible centering errors for the groups. We have tested the effect of centering errors by adding random offsets to the individual group centers, with r.m.s. values of $6\arcsec$, $12\arcsec$, and $24\arcsec$ in each coordinate. The resulting profiles are still well fit by our model, but 
the best-fit value of $r_s$ increases to 220, 260 and 340 \hkpc\ for the three cases respectively. This suggests average centering errors are $\lesssim 6\arcsec \sim$  
25--50 \hkpc\ in each coordinate, consistent with other estimates of the centering uncertainty (George et al.~in prep.).

We have also investigated other forms of stacking. In principle we could correct for the predicted variations in concentration, for instance, stacking in $r/r_s$, or we could stack in comoving rather than physical coordinates. Testing scaling the radius by $r_s$, $r_{vir}$, or $(1+z_L)$, we find little or no significant improvement in the $\chi^2$ of the fit to the radial profile. In particular, we find only a marginal 
indication of any trend in concentration with mass or redshift. Given that the halo-to-halo scatter 
measured in simulations is comparable or larger than to the average trends over the mass and 
redshift range spanned by our group sample, this is perhaps unsurprising. Furthermore, since the concentration relations are themselves dependent on cosmology, we would have to account for this dependence in our marginalization over cosmological parameters, so we will not attempt to correct for variations in concentration. We can also consider other analytic fits to the profile. We find that NFW is preferred over a cored isothermal profile at the 95\%\ confidence level, and preferred over a singular isothermal ($\Sigma(r) \propto r^{-1}$) profile at 97--98\%\ confidence. Thus, our stacked profile provides empirical confirmation of the NFW model, in agreement with other recent high-precision measurements of cluster density 
profiles \citep[][Okabe et al.~in prep.]{Umetsu}.

Finally, we note that the fit to the radial profile gives us an independent check of our empirical shear variance estimates. Because we fit the profile without binning, we have very large number of degrees of freedom and thus a narrow range of expected scatter in the reduced $\chi^2$. The best-fit NFW profile has a reduced $\chi^2$ of 0.931 with a very small (0.0027) expected scatter, so we conclude that our empirical variance is probably over-estimated by $\sim$ 7\%, corresponding to error bars which are 
3.6\%\ too big. We correct for this in all our subsequent analysis, multiplying the empirical shear variance 
by a factor of 0.931.

\section{Results}

\subsection{The Geometric Signal}

Given a functional form for the radial dependence of the surface mass density contrast $\Delta\Sigma(r)$, we can proceed to estimate $\Gamma(x)$ via Eqn.~\ref{eqn:gamma}. 
The weights in the sum can be calculated as the inverse variance of the $\Gamma_{ij}$
\begin{equation}
w_{ij} = ({\rm var}[\Gamma_{ij}])^{-1} = \left({\Delta\Sigma_i(r_{ij})}\over{\Sigma_{(c,\infty) i}}\right)^2{\rm var}[{\tilde{\gamma}}_{t,ij}]^{-1}\, .
\label{eqn:weights}
\end{equation}
where the variance of the tangential shear is determined empirically, as described in section \ref{subsec:shapes}.

The `model' here, the geometric sensitivity function $(1 - 1/x)$, is fixed, while the data vary as we change $\Sigma_0$ and $r_s$, which both change $\Delta\Sigma(r)$, and the cosmological parameters, which map the redshifts ($z_S, z_L$) onto $x$-values and also determine $\Sigma_{c,\infty}$. If we restrict ourselves to flat cosmologies with two components, matter and dark energy, then the goodness-of-fit depends on $\Sigma_0$, $r_s$, $\Omega_X$, and $w$. 

\begin{figure}
  \includegraphics[width=0.45\textwidth, keepaspectratio]{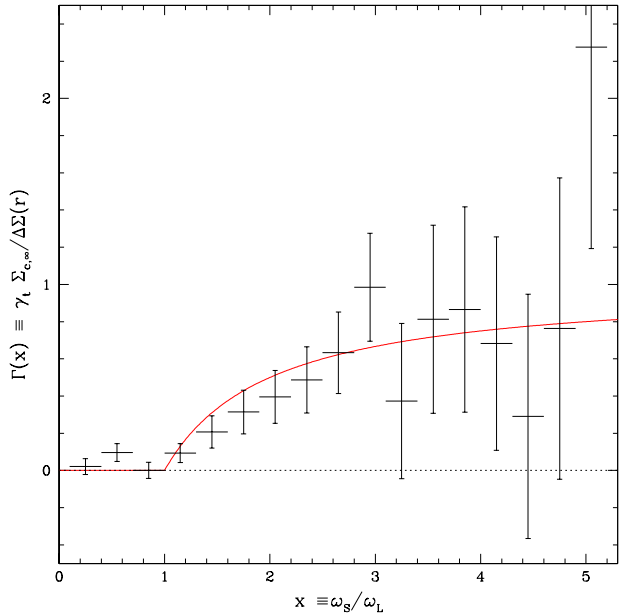}
  \caption{The geometric dependence of the lensing signal, plotted versus the distance ratio $x \equiv \omega_S/\omega_L$,  for a WMAP7 cosmology with $\Omega_M = 0.27$, $\Omega_\Lambda = 0.73$. The solid (red) curve shows the theoretical prediction $\Gamma(x) = 0$ for $x < 1$;  $\Gamma(x) = 1 - 1/x$ for $x > 1$.}
  \label{fig:gammax}
\vspace{0.1in}
\end{figure}

Fig.~\ref{fig:gammax} shows $\Gamma(x)$ over the range $x = [0,5]$ for a WMAP7 cosmology with 
$\Omega_M = 0.27$, $\Omega_\Lambda = 0.73$. The points are weighted averages in bins of 
0.3 in $x$, while the solid (red) curve is the theoretical expectation:
\begin{eqnarray}
 \Gamma(x) &=& 0 {\rm \ \ \ \ \ \ \ \ \ \ \ for\ \ }  x < 1\,; \\ \nonumber
 		   &=& 1 - 1/x {\rm \ \ for\ \ } x \ge 1\,. 
 \end{eqnarray}
Weights here are inverse variance, as in Eqn.~\ref{eqn:weights}. The errorbars on the data points are calculated as usual for an inverse-variance weighted average:
\begin{equation}
\sigma_{\Gamma, {\rm bin}} = \left(\sum_{\rm bin} w_{ij}\right)^{-1/2}\,,   
\end{equation}
where the sum is over all pairs $(i,j)$ with values of $x$ in the bin. 

Clearly the geometric signal is present in the COSMOS data, and measured to reasonable significance over a broad range of distance ratio $x$. Given the possible systematics in the measurement discussed below, the excellent agreement between theory and data illustrates the potential of the method. On the other hand, ``$\chi^2$-by-eye'' is somewhat misleading for this figure, as the binning in $x$ may hide systematics at particular distance ratios. 
(There is a $1.9\sigma$ indication of positive signal in one bin 
at $x < 1$, for instance. This could indicate photo-z errors 
are scattering sources to lower redshifts, but given the 
number of bins there is a 40\%\ chance that this is simply 
a random statistical fluctuation.) 
As discussed below, in a small field structures at a few redshifts can dominate the lensing signal, introducing excesses or deficits of mass along the line of sight that dominate the signal at particular values of $x$. The bins chosen here are broad enough to smooth out many of these features, but clearly a goodness-of-fit measurement over the whole data set is required to determine the statistical significance of the apparent agreement in Fig.~3. 

\subsection{Parameter Constraints}\label{subsec:paramconstraints}

We can estimate goodness-of-fit by calculating
\begin{equation}
\chi^2 = \sum_{ij}w_{ij}[(1 - 1/x_{ij}) - \Gamma_{ij}]^2
\label{eqn:chigamma}
\end{equation}
This sum depends on $\Sigma_0$, $r_s$ and the cosmological parameters, so marginalizing over the first two parameters gives constraints on the dark energy density $\Omega_X$ and the equation-of-state parameter $w$. While the resulting value of $\chi^2$ will tell us whether the data are a good fit to the model, this is not necessarily the most sensitive way of determining cosmological parameters. In particular, for flat $(k = 0)$ cosmological models with a cosmological constant with density parameter $\Omega_X = \Omega_\Lambda$, as the value of  $\Omega_\Lambda$ increases all distances will increase, and therefore so will $x$ and $(1 - 1/x)$. Thus the signal from a given source-lens redshift pair will be compared to $\Gamma(x)$ at a value of $x$ which is larger by some factor. The most sensitive probes of this re-scaling will be points at large $x$. To constrain $\Omega_X$ more precisely, individual measurements should be weighted by this sensitivity. 

\begin{figure}
\vspace{0.3in}
  \includegraphics[width=1.0\columnwidth, keepaspectratio]{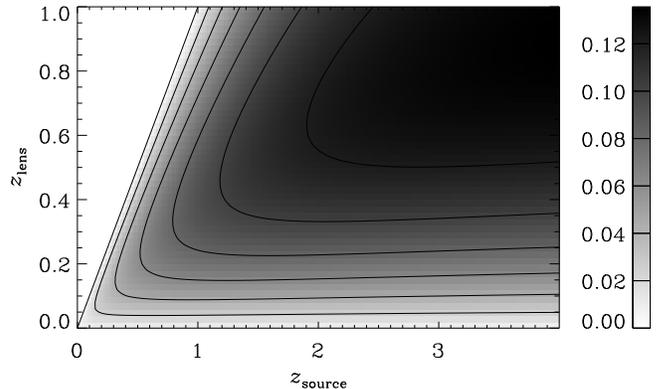}
  \caption{The sensitivity function $\Delta\Gamma(z_S, z_L)$. Points with high source or lens redshifts are the most sensitive to changes in cosmology.}
  \label{fig:lweight}
\end{figure}

The exact sensitivity to cosmology itself depends on the cosmological parameters. We can estimate a sensitivity factor and thus a weighting that will be close to optimal over the whole range of $\Omega_\Lambda$, however, by calculating 
\begin{eqnarray}
\Delta\Gamma(z_S, z_L) &=& \Gamma_{01}[x(z_S,z_L)] -  \Gamma_{09}[x(z_S,z_L)]\nonumber\\
&=& [x_{09}(z_S,z_L)]^{-1} - [x_{01}(z_S,z_L)]^{-1}
\end{eqnarray} 
where $ \Gamma_{01}$ $\Gamma_{09}$ are the model $\Gamma(x)$ evaluated for cosmologies with 
$\Omega_\Lambda = 0.1$ and 0.9 respectively (with $w = -1$ in each case).
Fig.~\ref{fig:lweight} shows this sensitivity function versus $(z_S, z_L)$. (Note that this sensitivity function has been calculated previously, e.g.~in Figure 3 of \citealt{Golse02}). For $z_S \le z_L$, the sensitivity function is zero since the model value $\Gamma(x) = 0$ independent of cosmology. For $z_S \ge z_L$, sensitivity generally increases with source or lens redshift.

\begin{figure}
  \includegraphics[width=0.45\textwidth, keepaspectratio]{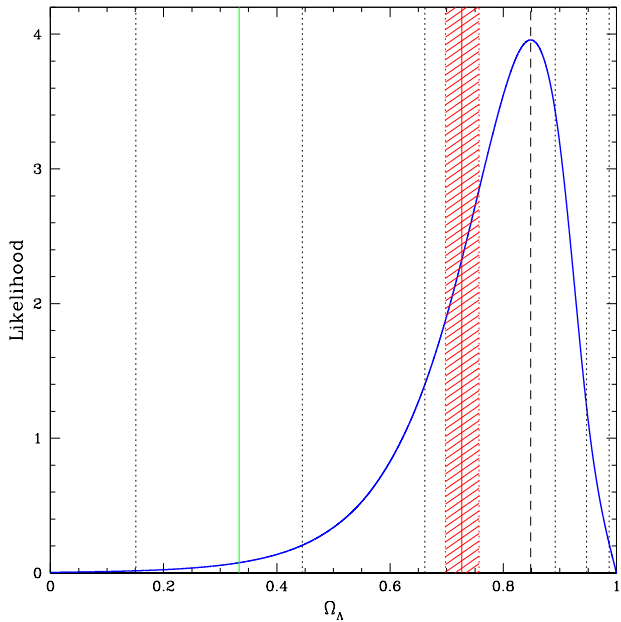}
  \caption{Likelihood function for  $\Omega_\Lambda$ with a uniform prior on $r_s$ over the range 120--200 \hkpc, using the restricted group sample. (The likelihood has been normalized so the area under the curve is 1.) The dashed vertical line indicates the value $\Omega_\Lambda = 0.848$ where the likelihood peaks. Dotted vertical lines indicate 68.2, 95.4 and 99.7\%\ (1, 2, 3-$\sigma$) confidence regions.
The solid vertical line and shading indicate the mean WMAP seven-year value 
$\Omega_\Lambda = 0.727$ and 68\%\ confidence range from \citet{WMAP7}. 
The vertical green line indicates the value $\Omega_\Lambda = 1/3$, the lower limit for cosmic acceleration ($q_0 < 0$).}
  \label{fig:omegax}
\end{figure}

To maximize signal-to-noise in our cosmological constraints, we apply this weighting quadratically in our previous expression for $\chi^2$:
\begin{equation}
\chi_w^2 = \sum_{ij}\Delta\Gamma^2_{ij}w_{ij}[(1 - 1/x_{ij}) - \Gamma_{ij}]^2/\sum_{ij}\Delta\Gamma^2_{ij}
\label{eqn:finalchi2}
\end{equation}
This $\chi^2$ can be converted to a likelihood by assuming the error distribution is Gaussian. This is
only approximately true in our case, but determining more accurate error distributions would require a significantly more complex error analysis, so we will leave this to future work. 
Fig.~\ref{fig:omegax} shows the likelihood function for  $\Omega_\Lambda$, calculated using the restricted group sample and normalized so the area under the curve is 1. The equation of state is fixed to $w = -1$ and we have marginalized over $\Sigma_0$ and $r_s$. The dashed vertical line indicates the value where the likelihood peaks. Dotted vertical lines indicate 68.2, 95.4 and 99.7\%\ (1, 2, and 3-$\sigma$) confidence regions. The solid vertical line and shading indicate the mean WMAP seven-year value $\Omega_\Lambda = 0.727$ and 68\%\ confidence range from \citet{WMAP7}. 

The results are insensitive to the priors in $\Sigma_0$ provided they are reasonably broad, but they depend strongly on the smallest values of $r_s$ considered. This is because of the dependence of inferred physical distance on cosmology. For low values of $\Omega_X$, the transverse distance inferred from a given angular separation on the sky is smaller. Conversely, if we fix $r_s$ to a small value, small values  of $\Omega_X$ are preferred. Because the radial variation of the profile over the range of our data is much stronger than the redshift variation, the radial fit drives the $\chi^2$ values and thus biases our cosmological results if unphysically small values of $r_s$ are allowed. The average scale radius is predicted to be $r_s \sim$ 160 \hkpc\ for our sample, based on theoretical concentration relations. Simulations show $\sim$50\%\ scatter in concentration from halo to halo \citep[e.g.][and references therein]{Zhao:2009, Reed}, but most of these variations should average out in the set of $\sim$100 objects considered here, provided they represent a reasonably unbiased sample. On the other hand the conversion from angular to physical radius will vary by $\sim$10--15\%\ for the range of cosmologies considered, and the range of uncertainty on our fitted value of $r_s$ is $\sim$ 130--210 \hkpc. Thus, we take a uniform prior on $r_s$ over the range 120--200 \hkpc, allowing for a variation of $\pm$25\%\ around the fiducial value $r_s$ = 160 \hkpc. Extending the range of our priors to higher values of $r_s$ has little effect on the results, while extending it to lower values of $r_s$ decreases the lower limit on $\Omega_X$. 

Overall, we obtain the estimate $\Omega_X (w = -1) = \Omega_\Lambda = 0.848^{+0.0435}_{-0.187}$, corresponding to a detection of dark energy at more than 99\%\ confidence. This value is consistent with the most recent WMAP analysis of CMB anisotropies, which finds a mean value $0.727^{+0.030}_{-0.029}$ \citep{WMAP7}. Our 68\%\ confidence range is approximately 4 times wider than that of WMAP; given the small size of the field considered here (1.64 deg$^2$), however, this level of precision demonstrates the power of the geometric test. We note however that our error estimates do not include systematic effects. We estimate the magnitude of some of the possible systematics in the next section.

In a cosmology with multiple components with equation-of-state parameters $w$ and 
density parameters $\Omega_w$ , cosmic dynamics can be characterized by the 
deceleration parameter
\begin{equation}
q = {1\over 2}\sum_w \Omega_{w} (1+3w)\, .
\end{equation}
 For a flat universe with components $\Omega_m$ and $\Omega_\Lambda$, we find a value $q_0 = -0.77^{+0.28}_{-0.066}$ at the present day. 
 Present-day acceleration, which corresponds to $q_0 < 0$ or 
 $\Omega_\Lambda > 1/3$ if $w=-1$,  is detected at greater than the 98\%\ C.L.~(solid [green] vertical line in Fig.~\ref{fig:omegax}).

\begin{figure}
 \vspace{0.2in}
\includegraphics[width=0.45\textwidth, keepaspectratio]{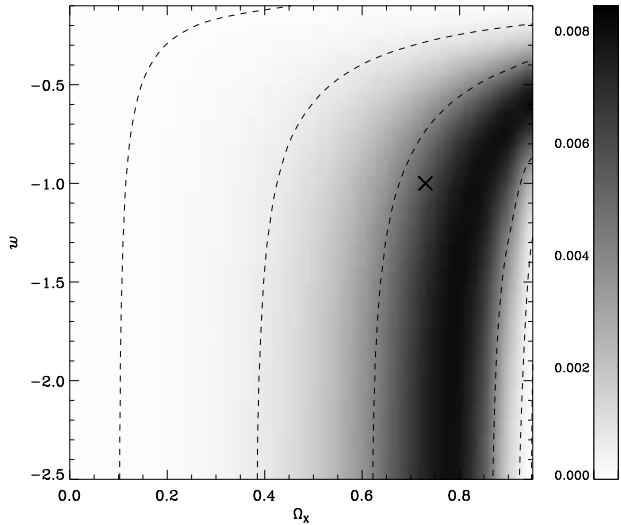}
  \caption{Likelihood surface for the cosmological parameters $\Omega_X$ and $w$ (grayscale). The dotted contours indicate regions enclosing 68.2\%, 95.4\%\ and 99.7\%\ (1, 2, 3-$\sigma$) of the probability in the region covered by our priors ($-2.5 \leq w \le -0.1$, $0 \leq \Omega_x \leq 1$). The ``X'' marks the mean WMAP seven-year value from  \citet{WMAP7}
}
  \label{fig:womegax}
\end{figure}

We can also extend the constraints to more general dark energy models with $w \neq -1$. Fig.~\ref{fig:womegax} shows the likelihood surface for models with a dark energy component $\Omega_X$ with an equation-of-state parameter $w$. Our current results provide an upper bound of $w < -0.4$ (68.2\%\ C.L.). They do not provide a lower bound, although there is some information in the constraints on $w$ as a function of $\Omega_\Lambda$. The shape of the confidence regions is similar to those derived by geometric tests using strong lensing \citep[e.g.][]{Jullo10}, although our contours are shifted upwards to less negative values of $w$, perhaps because of the redshift distribution of our lenses. The shape of the confidence regions also differs from those derived from observations of the CMB  \citep{WMAP7}, supernovae \citep{Sullivan11}, or BAO \citep{Percival10, Beutler11}, providing interesting complementarity with these other methods. 

\subsection{Cosmic Sample Variance and Systematics}\label{subsec:4.3}

\subsubsection{Basic Noise Sources}

There are several basic sources of noise in our measurement of $\Gamma(x)$. Individual galaxy 
shapes are noisy estimators of the tangential shear, and shape measurements themselves are not perfect, particularly for small or faint galaxies. The redshift errors in our lens population are negligible 
($\sim$ 0.0017 on average), but the photometric redshift errors for the source population may contribute significantly to the errors, once again particularly for faint galaxies at high redshift. These statistical errors are included, at least approximately, in our estimates of $\chi^2$. Our likelihood estimator also assumes a particular form for the average radial density profile of the groups $\Sigma_0$. This profile may not correspond exactly to the true mean surface mass density contrast, and individual groups may deviate from the average in ways that depend systematically on redshift. The tests described in section \ref{subsec:radial} suggest these effects are at or below the level of our statistical errors.

The relative importance of the source or lens sample selection is less clear a priori. To explore the dependence of our results on source galaxy properties we have rerun parameter constraints with samples cut in magnitude range, S/N, size, and redshift error. In each case reducing the sample size increases the final error in $\Omega_\Lambda$, but there is no obvious systematic trend for cuts in magnitude, S/N or size, beyond the dependence on source redshift discussed in section \ref{subsubsec:cosmicvar} 
below. The precision of the final results is quite dependent on the redshift cuts we make in the source sample and on cuts to the group sample; including all the groups in the full (rather than the restricted) sample increases the uncertainties by $\sim$25\%, including source galaxies with {\sc zp\_sec$>0$} increases them by $\sim$30\%, and including sources with $\Delta z/(1+z) > 0.05$ doubles them. 
Examining the catalogue, it seems the particular sensitivity to redshift errors is related to sources at 
$z = $ 2--3, which contribute strongly to the constraints due to the sensitivity function (cf.~Fig.~\ref{fig:lweight}). Cutting these objects out, or diluting their signal with many poorly determined redshifts, both reduce out final sensitivity quite strongly. We have also experimented with using more restrictive cuts on redshift errors, or using  a larger aperture $10\arcmin$ aperture around each group; these do not improve our results significantly. 

Finally, the relative importance of redshift errors is also slightly unclear in our method. We have attempted to estimate the  contribution of photometric redshift errors to the total error budget by adding scatter to the photometric redshifts in the source catalogue. Adding scatter equal to $5\times$ or $10\times$ $\Delta z$ (increasing $\langle\Delta z/(1+z)\rangle$ to 0.09 or 0.18) increases our 68\%\ C.L. bounds on 
$\Omega_\Lambda$ by 12\%\ and 30\%\ respectively. This suggests that our current redshift errors contribute only 5--10\%\ of the total error budget. If so, this is a promising result for other surveys which are likely to have redshift errors larger than COSMOS.

\subsubsection{Cosmic Sample Variance -- Empirical Estimates}\label{subsubsec:cosmicvar} 

A more complicated source of systematic uncertainty in the measurement comes from structure along the line of sight. The COSMOS survey probes a relatively small field, and several large structures are clearly visible in the galaxy redshift distribution below redshift $z =1$. We can look for evidence of cosmic sample variance effects in various ways. One simple estimate of the importance of cosmic variance can be obtained by boot-strapping, splitting the sample into two or more disjoint sets. We can test for systematics by measuring how 
$\Delta_{68}$, the width of the 68\%\ confidence region for $\Omega_\Lambda$, changes relative to 
the Poissonian expectation $\Delta_{68} \propto \sqrt{n_c/n_0}$ when we cut the sample from 
$n_0$ sources down to $n_c$. Splitting the field into 4 quadrants with roughly equal numbers of source galaxies, we find $\Delta_{68}$ increases by a factor of 1.9--2.5, so this seems consistent with the factor $\sqrt{n_c/n_0} = 2$ expected from Poisson scaling. We note however that one quadrant gives a best fit value of $\Omega_\Lambda = 0$ (albeit with a 68\%\ uncertainty of $+0.6$) whereas the others give values of $\sim$0.9. Examining the lenses and source distribution in this quadrant, it is not immediately obvious whether specific structures produce this shift. This is a sobering lesson about the possible effects of cosmic sample variance. 

Splitting our lenses into two groups by redshift, each with $n_c \sim n_0/2$, we find $\Delta_{68}$ increases by a factor of 1.33 for $z_L > 0.4$ and 2.4 for $z_L < 0.4$. Thus the noise increases at a roughly Poissonian rate in the high redshift sample, while in the low redshift sample it increases much more quickly. Similarly in a sample cut at $z_S > 1$, $\Delta_{68}$ increases by 1.33, 
while for $z_S < 1$, it increases by 1.73. These results suggest the signal at low redshift is more prone to systematics. We have tested alternative weighting schemes that attempt to correct for the trend with source redshift. Down-weighting sources with $z_S < 1.5$ by a factor of 0.5 or 0.33 reduces $\Delta_{68}$ by 30\%\ or 40\%\ respectively, but moves the peak of the likelihood $1.3\sigma$ or $1.7\sigma$ away from the WMAP7 value. Thus, it seems there is some trade-off between precision and accuracy in the redshift weighting. Clearly this subject requires further theoretical work, using realistic simulations of large-scale structure. For the moment, in the absence 
of an optimal weighting scheme motivated by theory, we choose not to apply either weighting to our 
final results.

One final concern is that the average line of sight to all our groups could be slightly over- or underdense. This would introduce a baseline shift in our model of $\Gamma(x)$, modifying it to:
 \begin{eqnarray}
 \Gamma(x) &=& \ \Gamma_0 \phantom{(1 - 1/x) + } {\rm \ \ for\ \ }  x < 1\,; \nonumber \\
 		   &=& (1 - 1/x) + \Gamma_0 {\rm \ \ for\ \ } x \ge 1\,. 
 \end{eqnarray}
We can test for a constant offset $\Gamma_0 \ne 0$ most easily in the range $x = [0,1]$ where the expected signal is zero. We find $\Gamma_0 = 0.02 \pm 0.03$, so there is no significant evidence for an offset. Furthermore, adding $\Gamma_0 = \pm 0.02$ to our model gives essentially identical constraints on $\Omega_\Lambda$ (the peak value shifts by less than $0.1\sigma$), so the method appears to be robust to any small offset of this kind. 

\subsubsection{Cosmic Sample Variance -- Theoretical Prediction}

We can also use the error description of \citet{Taylor07} to estimate the effect 
on our measurements of structure along the line of sight. They calculate that the 
tangential shear induced by large-scale structure between the observer and two 
background galaxies at redshifts $z_i$ and $z_j$ introduces a covariance in shear 
measurements given by  
\begin{equation} 
{\rm var}[\gamma_{t,ij}]=\int_0^{\infty}\frac{\ell {\rm d} \ell}{2\pi}
C^{\gamma\gamma}_{ij}(\ell)\left(\frac{2[1-J_0(\ell\theta)]}{(\ell\theta)^2}-\frac{J_1(\ell\theta)}{(\ell\theta)}\right)^2 , 
\end{equation}
where we have integrated over a circular aperture of radius $\theta$ by multiplying by the Fourier transform of the aperture (the term in brackets), and 
$C^{\gamma\gamma}_{ij}(\ell)$ is the tomographic cosmic shear power
spectrum \citep{Hu:1999}. 

\begin{figure}
\centering
  \includegraphics[width=0.9\columnwidth]{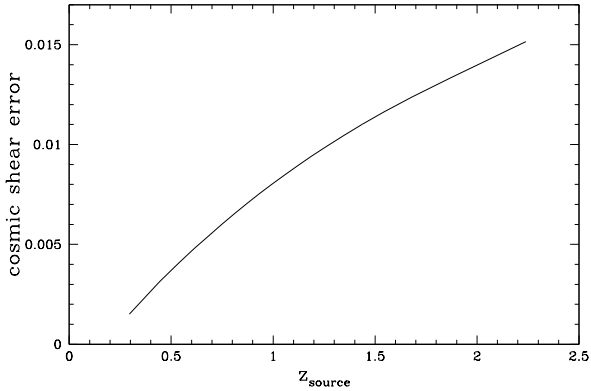}
 \caption{The predicted contribution to cosmic shear errors from unidentified structure along the line of sight up to a given source redshift. The contribution is integrated over a 6$\arcmin$ aperture, assuming a WMAP7 cosmology and the COSMOS source redshift distribution.}
 \label{fig:cs_plot}
\end{figure}

In our case, we are only concerned with the auto-correlations between redshift bins with $i=j$. 
These will give an estimate of the excess variance added to our shear measurements by cosmic 
structure, as a function of source redshift. Fig.~\ref{fig:cs_plot} shows this error term calculated for an aperture of 6$\arcmin$ in a WMAP7 cosmology and assuming the COSMOS redshift distribution.
We have calculated the shear error using $20$ discrete bins in redshift between $0<z\leq 2$. 
Ideally we would use a continuous cosmic shear in this measurement, as described in 
\citet{Kitching11}, but since the error contribution we find here is small and smoothly varying,
this approximation seems adequate. 

We see that the extra error term is always less than our empirical shear dispersion $\sigma_{\tilde{\gamma}} \ge 0.25$, and that it reaches a maximum of $\sim$6\%\ of the empirical dispersion. This suggests the contribution from cosmic shear is much smaller than the excess variance seen in the previous section, which may then be due to individual halos or to other systematics. Clearly more detailed simulations are needed to determine realistic cosmic sample variance errors for our particular technique, but these estimates reassure us that systematics do not completely dominate our current results. 

\section{Conclusions}\label{sec:conc}

Gravitational lensing is a versatile source of cosmological tests; it can be used to measure both the spectrum of matter fluctuations in the universe, and also its geometry. One particularly simple geometric test measures the rate at which the signal grows behind a lens and uses this to determine empirically the relationship between angular diameter distance and redshift. Here we introduce a new form of this test, the `stacked shear ratio test', which can be applied to large sets of individually weak lenses. By factoring out the source and redshift dependence from the expected amplitude of the lensing signal, we define a universal geometric dependence $\Gamma(x) \equiv (1 - 1/x)$, where $x \equiv \omega_S/\omega_L$ and $\omega_S, \omega_L$ are the comoving distances to the source and lens respectively, which can be estimated from individual measurements of tangential shear. As we vary cosmological parameters, the goodness-of-fit of our data with respect to the universal form provides a constraint on the geometry or equation of state of the universe. Physically, the test measures the relative distance scale behind the lenses; a higher value of $\Omega_X$ will increase this scale, so we can get particularly sensitive upper bounds on the density of dark energy.

The new method requires a significant lensing signal, although the signal can be spread over many individual lenses, and a large set of sources with well determined redshifts. The X-ray-selected group sample from the COSMOS survey provides a good test of the method, although cosmic sample variance is a concern over a field this small. Stacking groups with well-determined centers, we find a radial surface mass density contrast in good agreement with theoretical predictions. Examining the mean tangential shear behind the stacked group sample as a function of $x$, we see a clear manifestation of the geometric signal. 

Calculating goodness-of-fit for flat cosmological models with matter and a cosmological constant, with optimal weighting for sensitivity to cosmology over a broad range of $\Omega_X$, we obtain the constraints $ \Omega_\Lambda \equiv \Omega_X (w = -1) = 0.848 ^{+0.0435}_{-0.187}$ (68.2\%\ CL). Extended to dark energy models with constant equation-of-state parameters $w \neq -1$, our results do not constrain $w$ from below, but give the upper bound $w < -0.4$ (68.2\%\ CL). The shape of the confidence regions differs from those derived from observations of the CMB, supernovae or BAO, providing interesting complementarity with these other methods. Systematic uncertainties and the effect of cosmic sample variance are hard to estimate accurately without more detailed simulations, but the simple tests presented in Section \ref{subsec:4.3} suggest they do not dominate our statistical errors. Overall, the consistency of our results with more accurate determinations of $\Omega_\Lambda$ \citep{WMAP7} provides a good demonstration of the potential of the stacked shear ratio test in larger-area surveys. 

Our method does have several limitations. The simple scaling only applies for flat cosmologies; for 
$k \neq 0$ the geometric term in the lensing equation depends separately on $z_S$ and $z_L$, so a more complicated 2-D fit is required. We have also worked exclusively in the weak shear limit where $\kappa \sim \gamma \ll 1$. This does not restrict us significantly for COSMOS groups, since they are in the weak regime well into their central regions where other effects (e.g.~baryonic mass, centering errors) become important anyway. Applying our method to more massive clusters would require modeling the full shear, and thus introduce separate dependence on $z_S$ and $z_L$. We also 
assume a specific functional form for the surface mass 
density contrast, namely the projected NFW profile. In 
practice this form seems to be an excellent fit to the 
data, however, and therefore we would not expect our 
results to change significantly if we used instead of this 
an empirical profile based on the data alone.
Finally, our lower bound on $\Omega_X$ is sensitive to the average physical scale radius of the lenses; sensible priors and/or  reduced weighting at small radii are necessary here to avoid unphysical solutions which degrade the lower bound on $\Omega_X$.

For the large samples of relatively low-mass lenses such as those expected from 
current (e.g.~CFHTLenS\footnote[4]{planned site: http://www.cfhtlens.org}) and forthcoming lensing surveys (e.g.~DES\footnote[5]{http://www.darkenergysurvey.org}; PanStarrs\footnote[6]{http://pan-starrs.ifa.hawaii.edu}; LSST\footnote[7] {http://www.lsst.org}; HSC\footnote[8]{\citet{Takada10}}; 
Euclid\footnote[9]{http://sci.esa.int/euclid}; see \citealt{Massey10b} for a full listing), our method provides a simple, intuitive measure of geometry. 
The statistical uncertainties on $\Omega_\Lambda$ obtained for the 1.64 deg$^2$ COSMOS field are only $4\times$ larger than those obtained from current CMB results \citep{WMAP7}, and should scale as (survey area)$^{1/2}$. Thus if larger shear surveys can overcome systematics, there are good prospects for significantly reducing the uncertainties on abundance of dark energy. Given its sensitivity to sources at high redshift, our method might profitably be adapted  to lensing measurements based on magnification \citep{vanW10,Hildebrant11,Heavens11}. Since these require only measurements of magnitudes, not shapes, they may provide larger samples of galaxies at $z = 2$--3. It could also be combined with CMB lensing \citep{Hu:2007,Das:2009}, which would provide an anchor point at the largest observable values of $x$. In the longer term, as larger data sets become available and the sensitivity of lensing tests increases, their most interesting cosmological application might be to search for changes in the equation of state at particular redshifts. From this point of view, the method proposed here is flexible, as the redshift sensitivity depends simply on the distribution of source and lens redshifts used, and thus it can provide constraints over a range of redshifts, depending on the samples used. 

\acknowledgments
\noindent {\bf Acknowledgments}

The authors gratefully acknowledge helpful discussions with Niayesh Afshordi, Hendrik Hildebrandt, Mike Hudson, Fabian Schmidt, and Kris Sigurdson. JET is supported by a NSERC Canada Discovery Grant. RM is supported by STFC Advanced Fellowship \#PP/E006450/1 and ERC grant MIRG-CT-208994. AL acknowledges support from the Chamberlain Fellowship at LBNL and from the Berkeley Center for Cosmological Physics. 
The work of JR was carried out at the Jet Propulsion Laboratory, California Institute of Technology, under a contract with NASA. TK was supported by a RAS 2010 Fellowship. This work was supported by the World Premier International Research Center Initiative (WPI Initiative), MEXT, Japan.

The HST COSMOS Treasury program was supported through NASA
grant HST-GO-09822. We wish to thank Tony Roman, Denise Taylor, and
David Soderblom for their assistance in planning and scheduling of the
extensive COSMOS observations.  We gratefully acknowledge the
contributions of the entire COSMOS collaboration consisting of more
than 70 scientists.  More information on the COSMOS survey is
available at {\url{http://cosmos.astro.caltech.edu}}. It is also a
pleasure the acknowledge the excellent services provided by the NASA
IPAC/IRSA staff (Anastasia Laity, Anastasia Alexov, Bruce Berriman and
John Good) in providing online archive and server capabilities for the
COSMOS datasets.

\pagebreak

\bibliographystyle{apj}

\end{document}